# Strangeness Enhancement and System Size in the Hadronic Gas Model[1]

Azwinndini Muronga and Jean Cleymans
Department of Physics, University of Cape Town
Rondebosch 7700, South Africa

## Abstract

Strange particle enhancement in relativistic ion collisions is discussed with particular attention to the dependence on the size of the volume and/or the baryon number of the system.

## 1 The Hadronic Gas Model.

The motivation for using a "fireball" description for hadronic interactions comes from the work of Hagedorn [1]. In such a model one assumes that a thermal system is produced. This thermal "fireball" then expands until it freezes out, with the hadronic resonances decaying into the lightest particles. Hagedorn pointed out that the production of heavy particles in high energy proton-proton collisions calls for the canonical ensemble formalism instead of the grand canonical formalism due to the small interaction volume and the small number of particles. This basic idea has been developed by many authors [2] [3] [4] [5]. It has been shown [6] that the two formalisms will approach each other at large values of $B$ or volume. It has been shown in particular that if the radius of the volume $R \leq 4 fm$ and the baryon number $B$ is less than 30 one should use the canonical formalism.

In the grand canonical formalism the partition function of the hadronic gas in thermal and chemical equilibrium is given by

$$\ln Z(T, \mu_B, \mu_S) = \sum_i [W_i^m + (\lambda_B^{B_i}\lambda_S^{-S_i} + \lambda_B^{-B_i}\lambda_S^{S_i})W_i] \ . \tag{1}$$

---

[1] Talk presented by A. Muronga at RHIC'97 Summer Study 6-16 July 1997. A Theory Workshop on Relativistic Heavy Ion Collisions.

Here the first term refers to non-strange mesons and the second term to particles which carry baryon numbers $B_i$ and strangeness $S_i$. The fugacities are related to the baryon and strangeness chemical potentials $\lambda_B \equiv \exp(\mu_B/T)$ and $\lambda_S \equiv \exp(\mu_S/T)$. $W_i$ is the phase space factor for hadrons of species $i$ (mesons, baryons and their antiparticles). The phase space factors are given by

$$W_i = \frac{d_i m_i^2 VT}{2\pi^2} K_2(m_i/T) \ , \qquad (2)$$

with $d_i$ denoting the spin degeneracy, $m_i$ the mass of hadron species $i$, $V$ the volume of the system and $K_2$ is the Bessel function of the second type. The thermal contribution of the particle multiplicity $N_i = W_i$ has to be added by resonance contributions to get the final particle multiplicity

$$N_i = W_i + \sum_j Br(j \longrightarrow i) W_j \ . \qquad (3)$$

Here $Br(j \longrightarrow i)$ is the branching ratio of the decay of resonance $j$ to particle $i$. For a final state which is in both thermal *and* chemical equilibrium, all hadronic ratios should be determined by the values of $T$ and $\mu_B$ which can be fixed using two ratios. In this work, we use the canonical formalism and our results will always be compared with the grand canonical results. In the canonical formalism, the exact treatment of quantum numbers is obtained by projecting the partition function onto the desired values of the conserved quantum numbers e.g $B$ and $S$. The partition function is then given by

$$Z_{B,S} = \frac{1}{2\pi} \int_0^{2\pi} d\phi \, e^{-iB\phi} \frac{1}{2\pi} \int_0^{2\pi} d\psi e^{-iS\psi} Z(T, \lambda_B, \lambda_S) \qquad (4)$$

where the fugacities $\lambda_B$ and $\lambda_S$ have been replaced by

$$\lambda_B = e^{i\phi} \qquad \lambda_S = e^{i\psi}. \qquad (5)$$

To apply this formalism it is useful to group all particles in the Particle Data Booklet [8] according to their quantum numbers (we leave out charm and bottom). We do not include cascade particles (the $\Xi$'s and $\Omega$'s) as their contribution is unimportant for the energy range under consideration, however it can be done [6]. We consider now the behavior at freeze-out. In this case all the resonances in the gas are allowed to decay into the lighter stable particles. This means that each particle density is multiplied by its appropriate branching ratio (indicated by $Br$ below). The abundances of particles in the final state are thus determined by [9]

$$n_H = \sum n_i Br(i \to H), \qquad (6)$$

where the sum runs over all particles contained in the hadronic gas and $H$ refers to a hadron ($\pi^\pm, K^\pm, ...$). Here $n_i$ is given by

$$n_i = \left[ Z_0 \frac{R_Q}{Z_{B,S}} \right] g_i \int \frac{d^3p}{(2\pi)^3} e^{-E_i/T} \qquad (7)$$

in which the generating functions $R_Q$ are given by equation 7 of reference [9]. The factor in square brackets in the above equation replaces the fugacity in the usual grand canonical ensemble treatment. The results to be presented are obtained using this formalism (unless otherwise stated). The results are obtained for a fixed baryon density $B/V$, and we do not use chemical potentials $\mu_B$ and $\mu_S$ or $\mu_Q$ as done in [10] [11].

## 2  Kaon Yield.

Preliminary results on the dependence of hadronic ratios on the number of projectile participants have recently been presented by the E866 experiment (for E802 Collaboration) [13] for relativistic $Au - Au$ collisions at the BNL-AGS. We will use the fact that since the net baryon number, $B$ corresponds to the total number of participants, then $B = 2N_{pp}$, where $N_{pp}$ is the number of projectile participants with the factor 2 reflecting the symmetry of the $Au - Au$ collision system.

The hadronic ratios have been studied as a function of the net baryon number $B$ [9]. This will correspond to the investigation of the same ratios as a function of the total number of participants in the reaction. The study was done so as to incorporate large values of B. The dependence on $B/V$ and $T$ was investigated. It was found that at large values of $B$, the dependence of the ratios on $B$ becomes negligible.

Using the same formalism of the hadron gas model presented in this work, one can also analyze the E859 data presented by [12], particularly the kaon production. In figure 1 we compare our results to the experimental data. The model reproduces the quadratic dependence shown by the experimental data if the total number of participants is small. However, for large values of $B$ the quadratic dependence disappears and the kaon production increases linearly as a function of $B$. At large values of $B$ (which will corresponds to a large system like $Au - Au$) we have more $NN$ collisions which result in a greater kaon production. Thus the results of our Hadron Gas Model (HGM) behave exactly like what is seen in experiments. As one goes from $p - p$ to $A - A$ system one gets an increase in the kaon production which is exactly what we see as we go high in the baryon number $B$. Because in this HGM we fix the baryon density, $B/V$ ($n_B$), the volume is determined by the net baryon number. This means, therefore, that the dependence of the kaon production on $B$ implies a dependence on volume, $V$. Hence one should expect the same behaviour if one plots the kaon production as a function of $V$. The model also shows that the $Si + Al$ and the $Si + Au$ have slightly different freeze-out conditions, $T$ and $B/V$. We conclude that the model is in good agreement with the data.

## 3  The $K/\pi$ ratio.

In figure 2 we compare our results (of $K^+/\pi^+$) with the recent data from AGS [12][13]. As one can see, our results show a steep rise with the number of participants, $N_{pp}$, before levelling off while the experimental data indicates a slower rise than the model predicts. A good agreement is obtained with the results of the E866 collaboration [13]. The relevant temperature is around $T \approx 100\ MeV$, the baryon density is in the range of $B/V \approx 0.02 - 0.05/fm^3$, which indicates a considerable expansion before freeze-out. In the grand canonical ensemble this corresponds to a baryon chemical potential of $\mu_B \approx 540$ MeV.

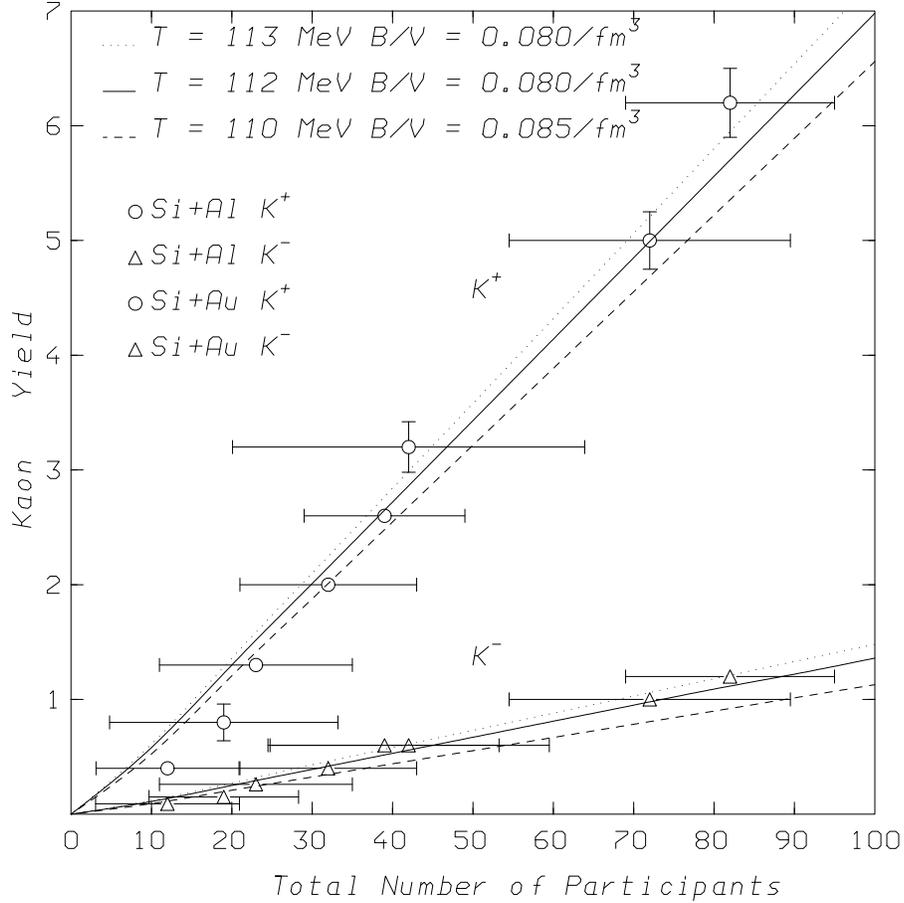

Figure 1: *The total yield of both positive and negative kaons is plotted as a function of the total number of participants in Si induced reactions at 14.6 GeV/c.*

The dependence of the $K/\pi$ ratio on the baryon density $B/V$ is shown in Figure 3. As one increases the baryon density, the baryon chemical potential $\mu_B$ increases correspondingly. Thus one expects the $K^+/\pi^+$ ratio to increase and the $K^-/\pi^-$ ratio to decrease. As one can see, the $K^+/\pi^+$ ratio initially rises with increasing baryon density, it then reaches a plateau and starts to decrease. For moderate densities, the $K^+/\pi^+$ ratio is almost independent of $B$, however, after reaching a maximum, the ratio becomes very strongly dependent on $B$. The $K^-/\pi^-$ ratio always decreases for increasing baryon densities. This decrease beyond $n_B = 0.1/\text{fm}^3$ is essentially caused by tilted baryon volumes (strong short-range repulsion between baryons). The two ratios are the same for a baryon free system, in agreement with $p\bar{p}$ collision data (vanishing baryon density) at CERN Intersecting Storage Rings (ISR) [14], [15], in which the $K^+/\pi^+$ and $K^-/\pi^-$ ratios were both 11% at midrapidity.

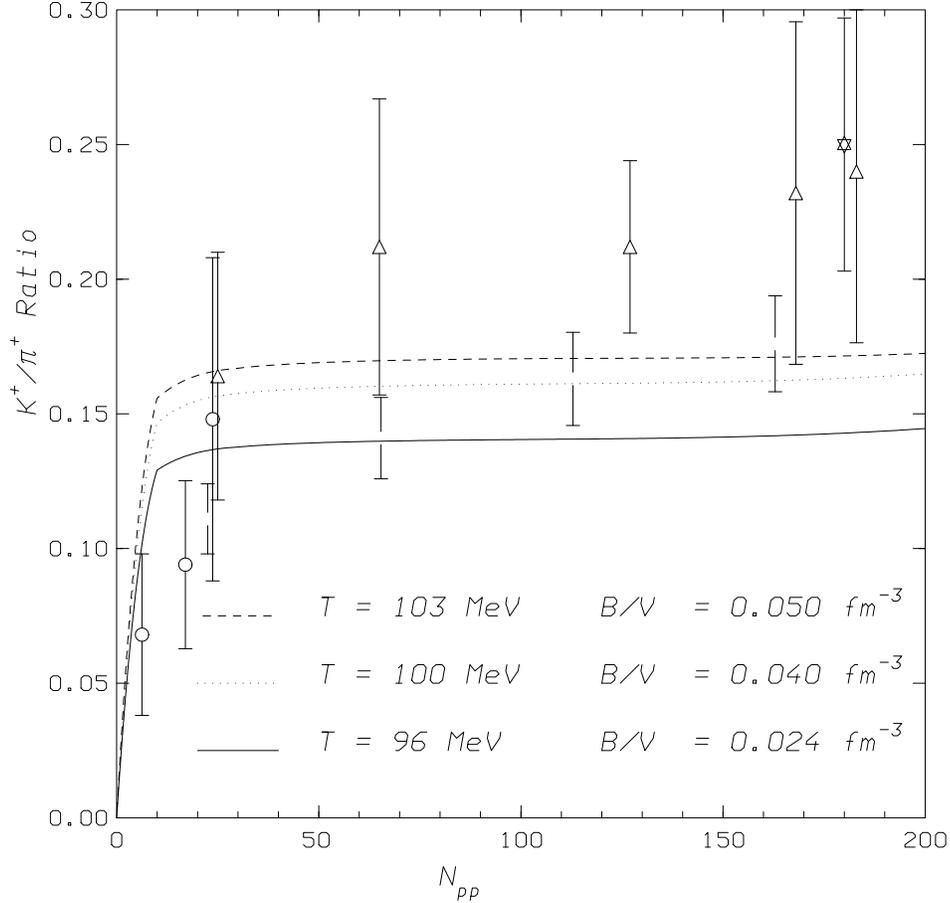

Figure 2: *The $K^+/\pi^+$ ratio as a function of the number of projectile participants, $N_{pp}$. The open circles indicate $Si - Al$ collisions.*

The dependence of $K/\pi$ ratio has been shown to have minimal or no dependence on the baryon density for large values of $B$ [9]. This is confirmed by calculations done in the grand canonical formalism [11].

Figure 4 shows the dependence of the $K/\pi$ ratios as a function of temperature. As the temperature increases, the kaon density and the pion density increase. However, the kaon density increases at a more rapid rate, thus one will expect the $K^+/\pi^+$ ratio to increase as $T$ increases. However in Figure 4, where we have fixed $B$, $B/V$ and obviously $V$, we see that the ratio will reach a maximum at a particular temperature depending on the realistic choices of $B$ and $B/V$, and as $T$ increases the $K^+/\pi^+$ ratio starts to decrease slowly while the $K^-/\pi^-$ ratio rises fast with temperature towards convergence with the $K^+/\pi^+$ ratio.

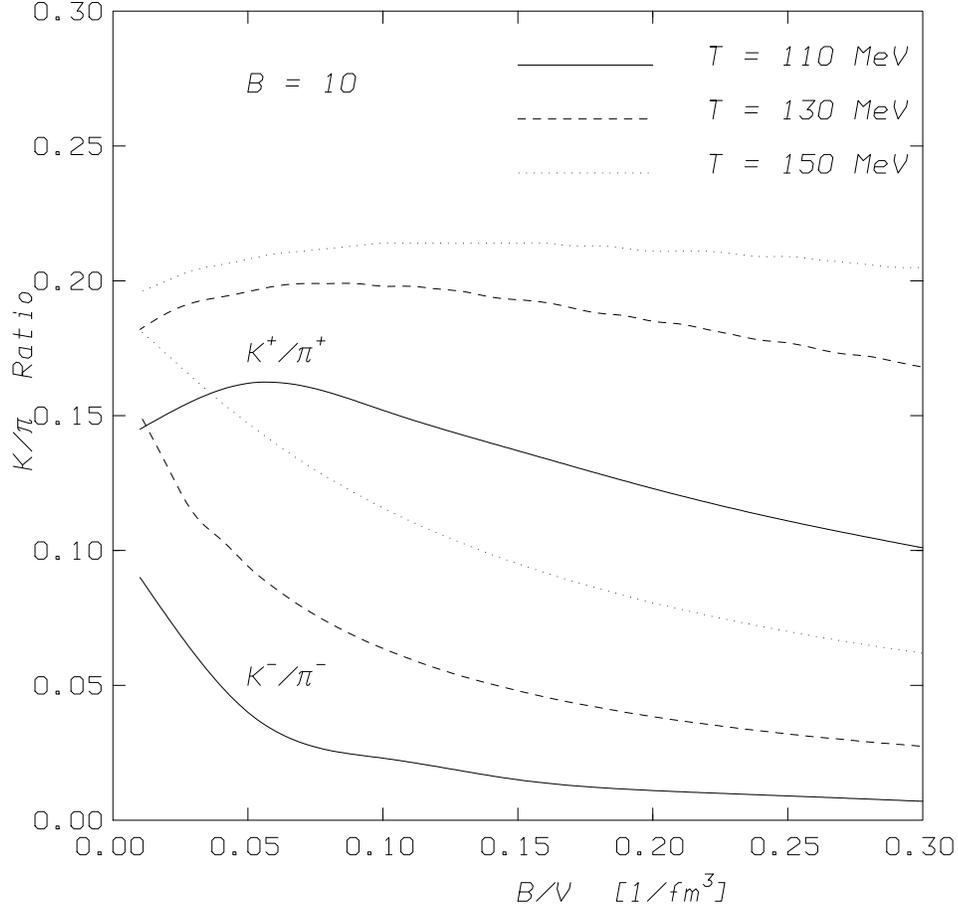

Figure 3: *The dependence of the $K/\pi$ ratios on the baryon density.*

At the hadron level one recalls that, as temperature increases, the probability of creating heavier resonances increases substantially, thereby increasing the number of decay channels, most of which favor pions. Thus we see that at large values of $T$ both ratios start to decrease.

## 4 The $K^+/K^-$ Ratio.

Another important ratio in the study of the Hadronic Gas Model is the $K^+/K^-$ which is shown in figure 5. The important feature here is the dependence of the ratio on $B$. For small values of $B$ the ratio increases very steeply before levelling off, and becomes almost independent of the volume for large values of $B$, remaining flat on average. This is so irrespective of the difference in the $K^+$ and $K^-$ yields. Thus one would expect that, at high values of $B$ where we will have more $K^-$'s absorbed due to large $K^-N$ inelastic cross-section, the ratio will be increased. However this is not seen in this ratio. This might be because as $B$ increases one produces $K^-$'s at a more rapid rate than $K^+$'s. When one combines this with the fact that in the region of large $B$ we also have large absorption of $K^-$'s, it leads to the $K^+/K^-$ ratio being insensitive to $B$ (and/or volume). For very large values of $B$ the ratio drops slightly. This is due to more $K^-$'s being produced with respect to $K^+$'s, even though one would expect a larger fraction of $K^-$'s to be absorbed in large systems because of the large $K^-N$ inelastic cross-section. This may make the $K^+/K^-$ ratio drop at large volumes. The $K^+/K^-$ ratio increases with increasing baryon density, with the increase being independent of $B$. The increase in the $K^+/K^-$ ratio as one increases the baryon density, $B/V$, is an indication that this ratio is a good probe for the baryon density of the gas.

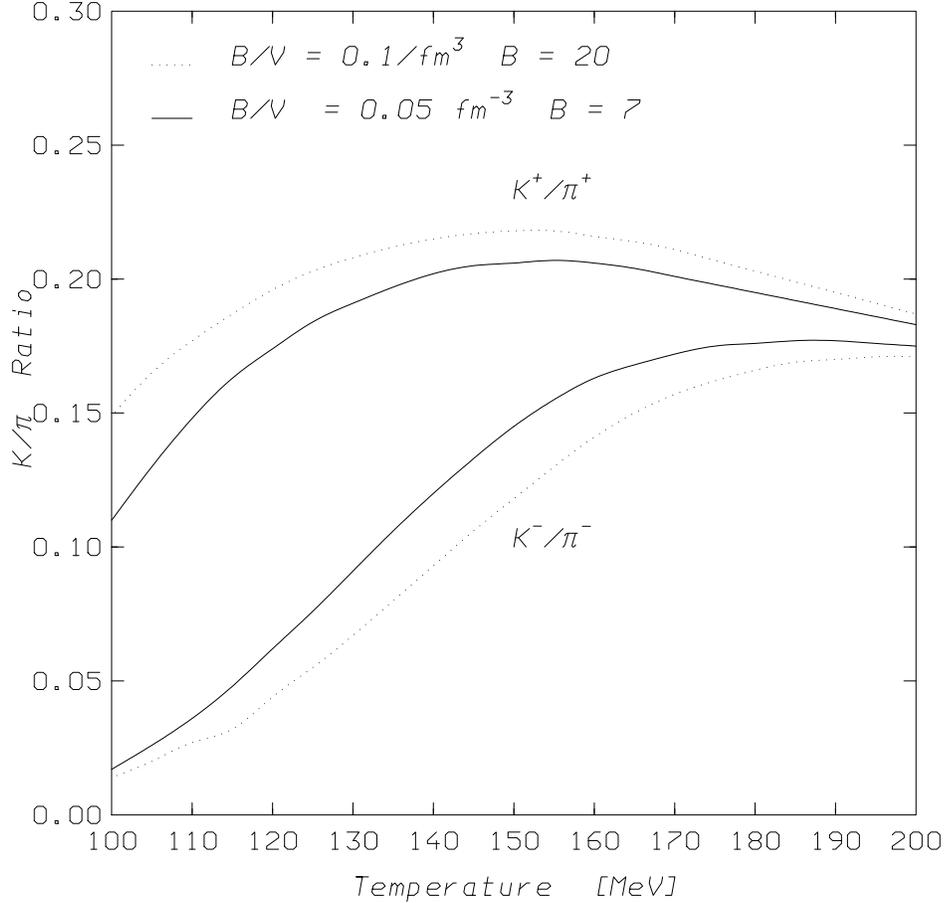

Figure 4: *The dependence of the $K/\pi$ ratios on the temperature.*

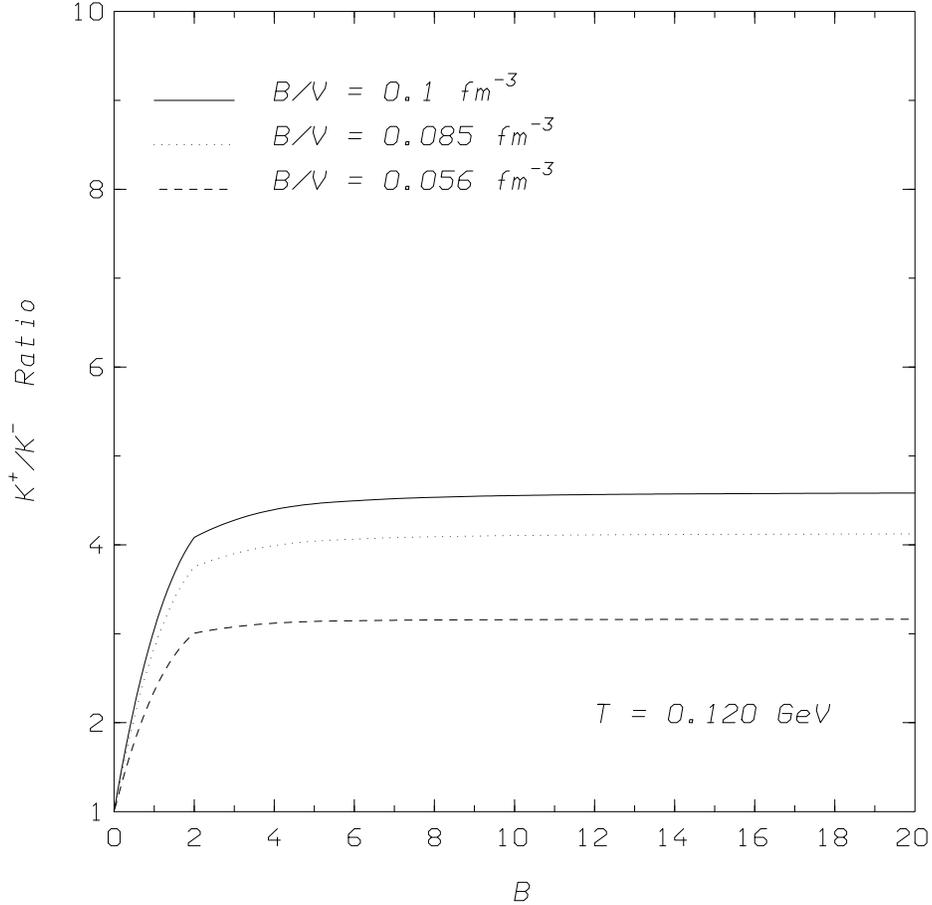

Figure 5: *The dependence of the $K^+/K^-$ ratio on the baryon density as a function of the baryon number $B$.*

## 5  Conclusion

The Hadron Gas Model has been compared with experimental results. The agreement is remarkable, keeping in mind that we used only one freeze-out temperature, $T$, and baryon density, $B/V$, for all particle species. Since particle emission is expected when the mean free path is comparable to the size of the system, different particles could decouple at different times which would imply different temperatures and baryon densities. In view of this one would not expect all the particle yields and the hadronic ratios to approach thermalization at the same time and the same rate. Also, the mechanisms which brought different particles towards freeze-out might be different. There are more processes in heavy ion collisions (such as $\pi - N$ and $\pi - \pi$ interactions, which lead to kaon production, than in nucleon-nucleon interactions. Thus one will expect an enhancement of the $K/\pi$ ratio. In view of this one should not take strangeness enhancement as the sole factor behind the $K/\pi$ ratio, but rather it should be taken as one of the ingredients in trying to understand the enhancement of the ratio. Also, because the pion production per nucleon decreases as one moves from $A-A$ via $p-A$ to $p-p$ collisions, one will expect the kaon/pion ratio to increase from $p-p$ to $p-A$ collisions until we have a constant kaon production. It should be noted that if it can somehow be shown that we do indeed reach thermalization in heavy ion collisions, that alone will not rule out the possibility of the claimed QGP's existence. One has to keep in mind that the thermal model provides only the scenario at the freeze-out and it does not tell us about the evolution and the dynamics of how the system got there.

# Acknowledgements

A.M. would like to thank the organizers of the RHIC Summer Study '97 for inviting him. He would also like to thank L. McLerran and B. Müller for their continuous support.

# References

[1] R. Hagedorn, CERN yellow report 71-12 (1971).

[2] B. H.-Th. Elze, W. Greiner and J. Rafelski, *Phys. Lett.* **B124** (1983) 515.

[3] R. Hagedorn and K. Redlich, *Z. Phys.* **C27** (1985) 541.

[4] J. Cleymans, E. Suhonen, G.M. Weber, *Z. Phys.* **C53** (1992) 485.

[5] J. Cleymans, K. Redlich and E. Suhonen, *Z. Phys.* **C51** (1991) 137.

[6] A. Muronga, M.Sc Thesis, University of Cape Town (1996) unpublished.

[7] J. Cleymans, K. Redlich, H. Satz and E. Suhonen, *Z. Phys.* **C58** (1993) 347.

[8] Review of Particle Properties, *Phys. Rev.* **D50** (1994) 1177.

[9] J. Cleymans and A. Muronga, *Phys. Lett.* **B388** (1996) 5.

[10] P. Braun-Munziger, J. Stachel, J.P. Wessels and N. Xu, *Phys. Lett.* **B344** (1995) 43.

[11] J. Cleymans, D. Elliott, H. Satz and R.L. Thews, *Z. f. Physik* **C74** (1997) 319.

[12] G.S.F.Stephans, In Proceedings of AIP Conference, "Strangeness in Hadronic Matter", J. Rafelski (Ed.), Tucson, AZ (1995) page 124.

[13] Y. Akiba for E802 Collaboration, Particle Production in Au+Au Collisions from BNL E866, Talk presented at Quark Matter '96, Heidelberg, Germany, may 1996.

[14] O. Villalobos Baillie, *Nucl. Phys.* **A525** (1991) 189c.

[15] G. Giacomelli and M. Jacob, *Phys. Rep.* **55** (1979) 1.